\documentstyle[12pt]{article}
\newcommand{\be}{\begin{eqnarray}}
\newcommand{\ee}{\end{eqnarray}}
\newcommand{\non}{\nonumber}

\begin{document}

\begin{titlepage}
\strut\hfill UMTG--195
\vspace{.5in}
\begin{center}

\LARGE  Simplified Calculation of Boundary $S$ Matrices \\[1.0in]
\large Anastasia Doikou, Luca Mezincescu and Rafael I. Nepomechie\\[0.8in]
\large Physics Department, P.O. Box 248046, University of Miami\\[0.2in]  
\large Coral Gables, FL 33124 USA\\

\end{center}

\vspace{.5in}

\begin{abstract}
The antiferromagnetic Heisenberg spin chain with $N$ spins has a 
sector with $N=$ odd, in which the number of excitations is odd.  In 
particular, there is a state with a {\it single} one-particle 
excitation.  We exploit this fact to give a simplified derivation of 
the boundary $S$ matrix for the open antiferromagnetic spin-${1 \over 2}$ 
Heisenberg spin chain with diagonal boundary magnetic fields.

\end{abstract}
\end{titlepage}

\section{Introduction}

The ground state of the antiferromagnetic spin-${1 \over 2}$ 
Heisenberg spin chain
\be
H= {1\over 4}\sum_{n=1}^N \left( \vec \sigma_n \cdot  \vec 
\sigma_{n+1} - 1 \right) \,, \qquad \qquad
\vec \sigma_{N+1} \equiv \vec \sigma_1 \,, \label{closed}
\ee 
lies in the ``even sector'' (i.e., $N=$ even). Most investigations of
the excitations of this model consider only this sector (see, e.g., Ref.  
\cite{FT1}), in which the number of excitations is necessarily even.

However, as remarked by Faddeev and Takhtajan \cite{FT2}, there is 
also the ``odd sector'' (i.e., $N=$ odd), in which the number of 
excitations is odd.  In particular, there is a state with a {\it 
single} one-particle excitation.

In this Letter, we exploit this fact to give a simplified derivation 
of the boundary $S$ matrix \cite{GZ} for the open antiferromagnetic 
spin-${1 \over 2}$ Heisenberg spin chain
\be
{\cal H} = {1\over 4}\left\{ \sum_{n=1}^{N-1} \vec \sigma_n \cdot 
\vec \sigma_{n+1} + {1\over \xi_-}\sigma^z_1 + {1\over \xi_+}\sigma^z_N 
\right\} \,, \label{open} 
\ee 
where the real parameters $\xi_\pm > {1\over 2}$ correspond to boundary 
magnetic fields.  Our analysis follows closely the one in \cite{GMN},
%which hereafter we refer to as I. 
which is a generalization of the approach developed by Korepin-Andrei-Destri
\cite{korepin}, \cite{andrei/destri} to calculate bulk two-particle $S$ matrices. 
However, in contrast to \cite{GMN}, we assume that $N$ is odd and 
consider one-particle states rather than two-particle states.  As a 
warm-up, we first briefly review the counting of excitations in both 
the even and odd sectors of the closed spin chain.

\section{Closed Chain Excitations}

In this section, we briefly review the enumeration of excitations of 
the closed spin chain, with Hamiltonian given by Eq.  (\ref{closed}).  
Upon adopting the ``string hypothesis'', the Bethe Ansatz equations 
lead to the following equations for the (real) centers 
$\lambda_\alpha^n$ of the strings (see, e.g., Refs.  \cite{FT1}, 
\cite{Nankai}):
\be
 h_n ( \lambda_\alpha^n ) = J_\alpha^n \,, 
\label{BAlog} 
\ee 
where $\alpha = 1, \cdots, M_n\,, \quad n = 1, \cdots, \infty \,.$
The so-called counting function $h_n (\lambda)$ is defined by
\be
h_n (\lambda) = {1\over 2\pi} \left\{ 
N q_n(\lambda) - \sum_{m=1}^\infty \sum_{\beta=1}^{M_m}  \Xi_{n m}
(\lambda - \lambda_\beta^m) \right\} \,, \label{hn}
\ee 
$\Xi_{n m} (\lambda)$ is given by
\be
 \Xi_{n m} (\lambda) = (1 - \delta_{n m}) q_{|n-m|}(\lambda) 
+ 2q_{|n-m|+2}(\lambda)
+ \cdots + 2q_{n+m-2}(\lambda) + q_{n+m}(\lambda) \,, \label{xi} 
\ee 
and $q_n (\lambda)$ is the odd monotonic-increasing function defined by
\be
q_n (\lambda) = \pi + i\log \left({\lambda + {in\over 2}\over \lambda 
- {in\over 2}} \right) 
\,, \qquad -\pi < q_n (\lambda) \le \pi \,. \label{q} 
\ee 
Moreover,
$\{ J_\alpha^n \}$ are integers or half-odd integers which satisfy 
\be
 -J_{max}^n \le J_\alpha^n \le J_{max}^n  \,,  \label{range1} 
\ee
where $J_{max}^n$ is given by
\be
 J_{max}^n =
{1\over 2}\left( N + M_n - 1 \right) - \sum_{m=1}^\infty min (m,n)\ M_m
\,. \label{jmax} 
\ee
We regard $\{ J_\alpha^n \}$ as ``quantum numbers'' which parametrize the 
Bethe Ansatz states.  For every set $\{ J_\alpha^n \}$ in the range 
given by Eq.  (\ref{range1}) (no two of which are identical), we assume 
that there is a unique solution $\{ \lambda_\alpha^n \}$ (no two of 
which are identical) of Eq.  (\ref{BAlog}).

The energy, momentum, and spin eigenvalues of the Bethe Ansatz states 
are given by
\be
E &=& - \pi \sum_{n=1}^\infty \sum_{\alpha=1}^{M_n} a_n (\lambda_\alpha^n)
\,, \label{energy} \\
P &=& - \sum_{n=1}^\infty \sum_{\alpha=1}^{M_n} \left[ q_n (\lambda_\alpha^n)
- \pi \right] \,, \label{momentum} \\
S &=& S^z = {N\over 2} - \sum_{n=1}^\infty n M_n  \,, \label{spin}
\ee 
where
\be
a_n(\lambda) = {1\over 2\pi} {d q_n (\lambda)\over d\lambda} 
= {1\over 2\pi} {n\over \lambda^2 + {n^2\over 4}} \,. 
\ee 

The number of holes (excitations) $\nu$ in a given Bethe Ansatz state 
is given by
\be
\nu &=& {\hbox{ number of vacancies for }} J^1_\alpha {\hbox{'s}} - 
{\hbox{ number of }} J^1_\alpha {\hbox{'s}} \non \\
&=& \left(J^1_{max} - J^1_{min} + 1 \right) - M_1 \,. \label{nu}
\ee 

The ground state lies in the even sector, with $M_1 = {N\over 2}$, 
and $M_n = 0$ for $n>1$.  We see from Eq.  (\ref{jmax}) that this 
state has $J^1_{max} = {N\over 4} - {1\over 2}$.  Therefore, the 
ground state has no holes; i.e., it is a ``filled Fermi sea'', with 
(see Eq.  (\ref{spin})) spin $S=S^z=0$.  Further calculations show that 
for $N \rightarrow \infty$, the energy and momentum of the ground state 
are given by
\be 
E_0 = - N \log 2\,, \qquad P_0={N \pi\over 2} \label{ground} \,.
\ee

Consider now the state in the odd sector with $M_1 = {N\over 2} - 
{1\over 2}$ and $M_n = 0$ for $n>1$.  This state has $J^1_{max} = 
{N\over 4} - {1\over 4} \,,$ and therefore it has 1 hole.  This state 
has spin $S=S^z={1\over 2}$.  Moreover, calculations along the lines 
of Refs.  \cite{FT1}, \cite{Nankai} show that the energy and momentum 
are given by
\be 
E = E_0 + \varepsilon(\tilde\lambda) \,, \qquad 
P = P_0 + p(\tilde\lambda) \,, \label{onehole} 
\ee 
where $E_0$ and $P_0$ are given by Eq. (\ref{ground}), 
\be 
\varepsilon(\lambda) = {\pi \over 2 \cosh \pi \lambda}  \,, \qquad 
p(\lambda) = \tan^{-1} \left(\sinh \pi \lambda \right) - {\pi\over 2} \,,
\label{energy/momentum} 
\ee 
and $\tilde\lambda$ corresponds to the hole rapidity.  This state 
consists of a single particle-like excitation (``kink'' or ``spinon'') 
with spin ${1\over 2}$, energy $\varepsilon(\tilde\lambda)$ and momentum 
$p(\tilde\lambda)$.  The energy-momentum dispersion relation is
\be
\varepsilon = -{\pi\over 2}\sin p  \label{dispersion} \,. 
\ee 

Similarly, one can show that there exist states with any 
(non-negative) integer number $\nu$ of excitations with the above 
dispersion relation.  States with $\nu=$ even lie in the even sector, 
while states with $\nu=$ odd lie in the odd sector.  The total energy 
and momentum are given by
\be 
E = E_0 + \sum_{\alpha=1}^{\nu}\varepsilon(\tilde\lambda_\alpha) \,, \qquad 
P = P_0 + \sum_{\alpha=1}^{\nu}p(\tilde\lambda_\alpha) \,. \label{manyhole} 
\ee 
Note that $E_0$ and $P_0$ (which are given by Eq.  (\ref{ground})) 
correspond to the ground-state energy and momentum only for $N=$ even.  
Indeed, for $N \rightarrow \infty$, the energy of a state with any 
finite number of excitations has an infinite contribution $E_0$ which 
must be subtracted.  In order to interpret the remaining finite part 
as the energy of the excitations, different subtractions must be 
performed in the even and odd sectors of the model.

\section{Open Chain Excitations}

We turn now to the open spin chain, with Hamiltonian ${\cal H}$ given by Eq.  
(\ref{open}).  The simultaneous eigenstates of ${\cal H}$ and $S^z$ have been 
determined by both the coordinate \cite{gaudin}, \cite{alcaraz} and 
algebraic \cite{sklyanin} Bethe Ansatz.

Using the string hypothesis, the Bethe Ansatz equations become \cite{GMN}
\be 
h_n (\lambda_\alpha^n) = J_\alpha^n \,, 
\ee 
where the counting function $h_n(\lambda)$ is now given by
\be 
h_n(\lambda) = {1\over 2\pi} \Big\{ (2N +1)q_n(\lambda) 
+ \sum_{l=1}^n \left[ q_{n+2\xi_+ - 2l} (\lambda)
                    + q_{n+2\xi_- - 2l} (\lambda) \right] \non \\
- \sum_{m=1}^\infty \sum_{\beta=1}^{M_m}\left[
  \Xi_{n m} (\lambda - \lambda_\beta^m)  
+ \Xi_{n m} (\lambda + \lambda_\beta^m) \right] \Big\}
\,. \label{counting/open} 
\ee 
The requirement that Bethe Ansatz solutions correspond to independent
Bethe Ansatz states leads to the restriction $\lambda_\alpha^n  > 0$.
The ``quantum numbers'' $\{J_\alpha^n\}$ are integers in the range
\be 
J_{min}^n \le J_\alpha^n \le J_{max}^n \,, \label{range2} 
\ee 
where
\be 
J_{max}^n - J_{min}^n = N + M_n - 2\left( \sum_{m=1}^\infty min (m,n)\ 
M_m \right) - 1 \,. 
\ee 
For simplicity, we assume that $2\xi_\pm$ is not an integer, and 
$\xi_\pm > {1\over 2}$.
 
The expressions for the energy and $S^{z}$ eigenvalues are the same 
as for the closed chain, namely, Eqs. (\ref{energy}), (\ref{spin}), 
respectively. (Of course, momentum and total spin are not good quantum 
numbers for the open chain.)

As in the case of the closed chain, the Bethe Ansatz state with a 
single one-particle excitation lies in the odd sector with $M_1 = 
{N\over 2} - {1\over 2}$ and $M_n = 0$ for $n>1$.  (The number of 
holes is again given by Eq.  (\ref{nu}).) This state has 
$S^z = {1\over 2}$.

We shall need in the next section the density $\sigma(\lambda)$ of 
roots and hole for this state, which is defined by
\be
\sigma(\lambda) = {1\over N} {d h_{1}(\lambda)\over d \lambda} \,.
\ee 
Passing with care from the sum in $h_{1}(\lambda)$ to an integral, 
we obtain an integral equation, whose solution is given by
\be
\sigma(\lambda) = 2 s(\lambda) + {1\over N} r^{(+)}(\lambda) \,, 
\label{sigma}
\ee
where
\be 
r^{(+)}(\lambda) =  s(\lambda) + J(\lambda) + J_+(\lambda) + J_-(\lambda) 
+ J(\lambda - \tilde\lambda) + J(\lambda + \tilde\lambda) \label{r}
\ee 
(plus terms that are higher order in $1/N$), $\tilde \lambda$ is the 
hole rapidity, and 
\be 
s(\lambda) &=& {1\over 2 \cosh \pi \lambda} 
= {1\over 2\pi} \int_{-\infty}^\infty d\omega\ e^{-i \omega \lambda}\
{e^{- |\omega|/2}\over 1 + e^{-|\omega|}} \,, \non \\
J(\lambda) &=& 
{1\over 2\pi} \int_{-\infty}^\infty d\omega\ e^{-i \omega \lambda}\  
{e^{- |\omega|}\over 1 + e^{-|\omega|}} \,, \non \\ 
J_\pm (\lambda) &=& 
{1\over 2\pi} \int_{-\infty}^\infty d\omega\ e^{-i \omega \lambda}\  
{e^{-(\xi_\pm - {1\over 2})|\omega|}\over 1 + e^{-|\omega|}} 
\label{definitions} \,. 
\ee

\section{Boundary $S$ Matrix}

The boundary $S$ matrix describes the interaction of an excitation 
with the ends of the spin chain.  The $U(1)$ symmetry of the 
Hamiltonian's boundary terms implies that the boundary $S$ matrix is 
of the diagonal form
\be
{\cal K}(\lambda \,, \xi) = 
\left( \begin{array}{ll}
      \alpha(\lambda \,, \xi) &0  \\
      0  & \beta(\lambda \,, \xi)  \end{array}\right) 
\,. \label{form} 
\ee 
We therefore need to explicitly determine the matrix elements 
$\alpha(\lambda \,, \xi)$ and $\beta(\lambda \,, \xi)$,
which are the boundary scattering amplitudes for excitations
with $S^z = +{1\over 2}$ and $S^z = -{1\over 2}$, respectively.

We proceed by examining states $| \tilde \lambda \rangle_a$ with a 
single one-particle excitation of rapidity $\tilde \lambda$.  The 
isotopic index $a$ is suppressed below.  For such states, the 
following simple quantization condition holds:
\be 
\left( e^{i 2 p(\tilde\lambda) N}\ 
{\cal K}(\tilde\lambda \,, \xi_-)\ 
{\cal K}(\tilde\lambda \,, \xi_+) 
- 1 \right) | \tilde \lambda \rangle = 0 \,. \label{quantization} 
\ee 
Here $p(\tilde\lambda)$ is defined by Eq.  (\ref{energy/momentum}), which is 
the expression for the momentum of a particle with rapidity $\tilde\lambda$ 
for the corresponding system with periodic boundary conditions.

For the $S^z = +{1\over 2}$ state, the quantization condition implies
\be
2 p(\tilde\lambda) + {1\over N} \Phi^{(+)} = {2 \pi\over N} m \,, 
\label{phase}
\ee
where $m$ is an integer and
\be
e^{i\Phi^{(+)}} = \alpha(\tilde\lambda \,, \xi_-)\ \alpha(\tilde\lambda \,, \xi_+)
\,. 
\ee  
On the other hand, one can derive the identity \cite{GMN}, \cite{Nankai} 
\be
2 p(\tilde\lambda) + {2\pi\over N}\int_0^{\tilde\lambda} 
r^{(+)}(\lambda)\ d\lambda + const
= {2\pi\over N} \tilde J \,, \label{show}
\ee 
where $r^{(+)}(\lambda)$ is the quantity introduced in Eq. (\ref{sigma}).
Comparing this result with Eq. (\ref{phase}), it follows that
\be
\Phi^{(+)} = 2\pi\int_0^{\tilde\lambda} r^{(+)}(\lambda)\ d\lambda
+ const \,.   
\ee 
Using the explicit expressions given in Eqs.  (\ref{r}), (\ref{definitions}) 
for $r^{(+)}(\lambda)$, we obtain the following result for 
$\alpha(\lambda \,, \xi)$ (up to a rapidity-independent phase factor):
\be
\alpha(\lambda \,, \xi) = 
{\Gamma \left({-i\lambda\over 2} + {1\over 4}\right) \over
  \Gamma \left({i\lambda\over 2} + {1\over 4}\right)} 
{\Gamma \left({i\lambda\over 2} + 1\right) \over
\Gamma \left({-i\lambda\over 2} + 1\right)}
{\Gamma \left({-i\lambda\over 2} + {1\over 4}(2\xi -1)\right)\over 
  \Gamma \left({i\lambda\over 2} + {1\over 4}(2\xi -1)\right)} 
{\Gamma \left({i\lambda\over 2} + {1\over 4}(2\xi +1)\right)\over
\Gamma \left({-i\lambda\over 2} + {1\over 4}(2\xi +1)\right)} \,.
\label{result1} 
\ee 

To determine the remaining element $\beta(\lambda \,, \xi)$ of the 
boundary $S$ matrix, we consider the $S^z = -{1\over 2}$ state.  The 
quantization condition implies
\be
2 p(\tilde\lambda) + {1\over N} \Phi^{(-)} = {2 \pi\over N} m 
\,, 
\ee 
with
\be 
e^{i\Phi^{(-)}} =\beta(\tilde\lambda \,, \xi_-)\ 
\beta(\tilde\lambda \,, \xi_+) \,. 
\ee 

The $S^z = -{1\over 2}$  state is most easily described within the Bethe Ansatz 
approach by changing the pseudovacuum to the state with all spins down.
Sklyanin has shown \cite{sklyanin} that there is a corresponding 
change $\xi_\pm \rightarrow -\xi_\pm$ in the Bethe Ansatz equations. 
The expression for the energy eigenvalues remains the same, but
the expression for the $S^z$ eigenvalues becomes
\be
S^z = \sum_{n=1}^\infty n M_n  - {N\over 2} \,.
\ee 
The $S^z = -{1\over 2}$ state now corresponds to the Bethe Ansatz 
state consisting of 1 hole in the Fermi sea ($M_1 = {N\over 2} - 
{1\over 2}$ and $M_n = 0$ for $n>1$). The calculation of the 
corresponding function $r^{(-)}(\lambda)$ is exactly the same as for 
$r^{(+)}(\lambda)$, except that we must track the change $\xi_\pm 
\rightarrow -\xi_\pm$.  We find that $\beta(\lambda\,, \xi)$ is given 
(up to a multiplicative constant) by
\be
\beta(\lambda\,, \xi) = -{\lambda +i(\xi -{1\over 2})\over 
\lambda -i(\xi -{1\over 2})} \alpha(\lambda \,, \xi) \,, 
\ee 
where $\alpha(\lambda\,, \xi)$ is given by Eq. (\ref{result1}).
This completes the derivation of the boundary $S$ matrix.

\section{Discussion}

We have seen that by considering one-particle states, the boundary $S$ 
matrix for the open Heisenberg chain can be obtained in a most direct 
and straightforward manner.  We expect that this approach can be used 
quite generally to calculate boundary $S$ matrices for integrable 
models with boundaries whose Bethe Ansatz equations are known, and in 
particular, for the models for which the method of Ref.  \cite{GMN} 
has already been successfully applied (see, e.g., \cite{essler} - 
\cite{tsuchiya}).  We are now using this approach to compute 
\cite{DMN} the boundary $S$ matrix for the anisotropic spin ${1\over 
2}$ chain, which was first calculated using the vertex operator method 
\cite{jimbo}.

\section{Acknowledgments}

This work was supported in part by the National Science Foundation 
under Grant PHY-9507829.

\vfill\eject

\end{document}